\documentclass[10pt]{article} 
\usepackage{amsmath,amssymb,cite,color,epsfig,graphicx,fullpage,hyperref}
\hypersetup{colorlinks=true,linkcolor=blue,citecolor=blue,urlcolor=blue}
\newcommand{\R}{{\mathcal R}}\newcommand{\bigO}{{\mathcal O}}

\begin{document}
\vspace*{10ex}
\begin{flushleft}
{\LARGE
\textbf{Reconstruction of disease transmission rates: \\[1ex]
applications to measles, dengue, and influenza}
}\\[6ex]
Alexander Lange\\[2ex]
Institute of Thermodynamics and Thermal Process Engineering, University of Stuttgart, Germany
\\[2ex]
E-mail: lange@itt.uni-stuttgart.de\\[4ex]
\end{flushleft}
\vspace*{1ex}
\subsection*{Abstract}
\vspace*{1ex}

Transmission rates are key in understanding the spread of infectious diseases. Using the framework of compartmental models, we introduce a simple method that enables us to reconstruct time series of transmission rates directly from incidence or disease-related mortality data.
The reconstruction exploits differential equations, which model the time evolution of infective stages and strains.
Being sensitive to initial values, the method produces asymptotically correct solutions.
The computations are fast, with time complexity being quadratic.
We apply the reconstruction to data of measles (England and Wales, 1948--67), dengue (Thailand, 1982--99), and influenza (U.S., 1910--27).
The Measles example offers comparison with earlier work.
Here we re-investigate reporting corrections, include and exclude demographic information.
The dengue example deals with the failure of vector-control measures in reducing dengue hemorrhagic fever (DHF) in Thailand.
Two competing mechanisms have been held responsible: strain interaction and demographic transitions.
Our reconstruction reveals that both explanations are possible, showing that the increase in DHF cases is consistent with decreasing transmission rates resulting from reduced vector counts.
The flu example focuses on the 1918/19 pandemic, examining the transmission rate evolution for an invading strain. Our analysis indicates that the pandemic strain could have circulated in the population for many months before the pandemic was initiated by an event of highly increased transmission.

\subsection*{Keywords} infectious disease modeling; disease transmission; differential equation models

\section{Introduction}

Essential for modeling an infectious disease epidemics is the knowledge of the transmission rates --- the rates at which susceptibles become infected by contagious individuals \cite{AM,GF08}.
These rates are determined by the contact behavior of the involved hosts as well as the risk of transmission during contact, they are specific to the pathogen and its transmission route \cite{LF09}. Transmission rates fluctuate and often systematically change over time as we will examine in this paper
for data of three different infections.

Combined with basic medical and demographic information, transmission rates are the most natural reference in predicting the time evolution of
disease prevalence.
Public health policies rely on such predictions, referring to the control of endemic infections as well as to the design of measures against emerging diseases, pandemics, or bio-terroristic threats \cite{FCC05,FDC09,LEH09}. 
In practice, however, there is no straightforward way these rates are obtained from epidemiological data.
The methods known are rather complicated and/or require lots of computing power \cite{FG00,BFG02,CF08,MF05,HERE11,WCBIL12,WYCIL13}, others are less developed 
at this stage
\cite{PWW11,H11,H12,M13}.
Addressing these issues, we will present a very simple method for reconstructing transmission rates from incidence or mortality data and illustrate its wide-ranging applicability.

There are lots of computing and modeling strategies in infectious disease epidemiology, including probabilistic simulations \cite{MF05,CF08}, networks \cite{RK03,KE05}, and compartmental models\cite{AM,KM27}, to mention a few. The conceptually simplest and mathematically most tractable ones fall into the last category.
In compartmental models,
transmissions between hosts are assumed to happen at random, incorporated through versions of the mass-action principle \cite{H05}.
Formulated in terms of differential equations, these models keep track of expectation values that represent
susceptible, infective, and recovered individuals (SIR), as well as other relevant compartments of the host population \cite{KM27}.
The transmission mechanism is implemented by a (time-dependent) coefficient, the transmission rate $\beta$.
Normalized with respect to an infection-specific removal rate $\lambda$,
the resulting unit-free number $\R_0=\beta/\lambda$ defines the basic reproduction of infections \cite{DHM90} --- a parameter well-known because of its importance for disease control and pathogen evolution \cite{AM,LF09}.

Whereas disease incidence and mortality define quantities of major interest to public health \cite{IIDDA}, transmission rates offer the more natural parameter in characterizing how a disease system is changing over time.
Due to the internal dynamics, disease prevalence often behaves strangely, for example, shows biennial time patterns even if infection is forced by an annual period (known for measles \cite{FC82,ERB00}) or increases even if risk factors are decreasing (reported for dengue \cite{TNS08}). Prevalence is sensitive to initial conditions and time resolution, it even shows chaotic behavior \cite{ERG98}.
The kind of information that can be captured by transmission rates
will be illustrated here for three examples: measles during the pre-vaccination era, the currently re-emerging dengue epidemic, and the Spanish flu pandemic.

The measles example is used to develop the method. It offers a test for the reconstruction results and its systematic errors. Data from England and Wales have been studied extensively, and much is known about the temporal pattern of transmissions \cite{FC82,BFG02,CF08}.

Epidemiological data are tainted with reporting errors, including missing or false diagnoses.
To compensate these errors the concept of the reporting proportion has been introduced (e.g., \cite{FC82,BFG02}).
In addition to the reporting proportion known from the literature,
we introduce a second one, enabling the definition of an effectively constant population, useful when demographic information is insufficient.

Besides focusing on technical questions --- regarding parameters, errors and extensibility to new compartments --- we study epidemiological questions and illustrate how the reconstruction can be applied to obtain conceptual results.
Here dengue represents a generic example.
Increasing cases of dengue hemorrhagic fever (DHF) in Thailand need to be explained based on decreasing transmission rates \cite{TNS08}.
Our method allows us to re-evaluate former explanations based on strain interaction \cite{NK08} and demographic transitions \cite{CIL08}.
The third application is the influenza pandemic 1918/19, which we study based on mortality data from the U.S. Here we use the reconstruction method to investigate irregularities prior to the pandemic peak, referred to as herald waves \cite{TM06}.

\section{Methods}

In this section we develop a procedure for reconstructing transmission rates
from time series data.
Measles infections are used as the main example. For other diseases the methodology requires adjustments, as we illustrate for two other infections in Results.

\subsection{Basic model}

One of the simplest settings for an infectious disease epidemic is given by a system of three first order differential equations,
\begin{subequations}\label{for040}
\begin{align}
\label{for041}S'&=\Gamma[S,I,R]-\beta\,\Sigma[S,I]\\
\label{for042}I'&=\beta\,\Sigma[S,I]-\lambda I\\
\label{for043}R'&=\gamma I-\rho R\,.
\end{align}
\end{subequations}
These equations determine the time evolution of compartments, $S,I$, and $R$, which in respective order quantify the average numbers of {\em susceptible, infective}, and {\em recovered} individuals \cite{KM27}.
Their sum,
\begin{align}\label{for037}
N=S+I+R\,,
\end{align}
represents the size of the {\em host population}.
Usually, this number differs from population data published in demographic reports (e.g., \cite{HA99}); it only includes parts of the population. Most obviously this applies to sexually transmitted infections where the host population represents the sexually active part.

Utilized as balance equation, $N'=\nu-\mu N$, the derivative of \eqref{for037}
determines the rate,
\begin{align}\label{for038}
\Gamma=\nu-\mu S+\delta R+\Theta[I,R]\,,
\end{align}
at which individuals enter, leave, and (possibly) re-enter the susceptible compartment --- through birth $\nu$, death $\mu S$, 
and (e.g., for flu-like infections)
after recovery $\delta R$. 
The remaining term,
\begin{align}\label{for039}
\Theta=(\lambda-\gamma-\mu)I+(\rho-\delta-\mu)R\approx0\,,
\end{align}
obtained by adding the three equations \eqref{for040} and comparing the result to \eqref{for038},
incorporates disease-induced mortality rates specific to the compartments $I$ and $R$.
For many epidemics these rates can be neglected. The recovery rate and the decay of immunity are then given by $\gamma=\lambda-\mu$ and $\delta=\rho-\mu$, respectively.

The number of transmissions is determined by the rate at which susceptibles become infected. This rate, $\beta\Sigma$, which defines removal in \eqref{for041} and incidence in \eqref{for042}, involves a functional and a coefficient. The functional, $(S,I)\mapsto\Sigma[S,I]$, represents the contacts between infected and susceptible individuals; usually it is modeled via mass-action \cite{H05},
\begin{align}\label{for004}
 \Sigma=SI/N\,.
\end{align}
Refined models contain fractional powers of $S$ and $I$ \cite{LHL87}.
The coefficient, $\beta$,
defines the {\em transmission rate} ---
the parameter we intend to reconstruct from time series data.

\subsection{Beta reconstruction}

The starting point is a version of \eqref{for042},
\begin{align}\label{for003}
I'=W-M/q\,,
\end{align}
which models the time evolution of disease prevalence in terms of quantities that are observable and thus stored in epidemiological data bases (e.g., \cite{IIDDA}).
$W$, given by $W=\beta\Sigma$, represents incidences of infection (number of new infections per time unit). $M$, given by $M=q\lambda I$, represents incidences related to a focal condition that infected individuals are affected by; the parameter $q$ models the associated likelihood.
If, for example, the infection causes deaths then $M$ can be utilized to model the induced mortality. Here the case-fatality ratio $c$ determines the likelihood, $q=\frac{1}{c}\frac{\lambda-\mu}{\lambda}\approx\frac{1}{c}$. Another example for such a condition is {\em being recovered}\,. Due to \eqref{for042} and \eqref{for043}, recovery from infection happens with likelihood $q=\gamma/\lambda$ at rate $M=\gamma I$.

By comparing the incidence terms in \eqref{for003} and \eqref{for042} we obtain a formula for the transmission rate,
\begin{align}\label{for009}
\beta(t)=\frac{W(t)}{\Sigma[S(t),I(t)]}\,.
\end{align}
Incidences $W$ are observed but contacts $\Sigma$ need to be calculated.
To determine $S(t)$ and $I(t)$
we distinguish two cases, proposing that either incidence $W(t)$ or mortality $M(t)$ is given. For now, we assume that these data are smooth functions over time.

{\em First case:} $W(t)$ is given; $I(t)$ and $S(t)$ need to be calculated.
The time-evolution of the infectives is reconstructed by solving a first order linear differential equation,
\begin{align}\label{for005}
I'+\lambda I=W\,,
\end{align}
which combines \eqref{for042} and \eqref{for009}.
The solution (see, e.g., \cite{BD05}),
\begin{subequations}\label{for00252a}
\begin{align}\label{for0052}
I(t)=I(0)e^{-\lambda t}+\int_0^te^{-\lambda(t-r)}W(r)\,dr\,,
\end{align}
involves a convolution integral and an initial value, which without additional information we approximate by the steady state solution, $I(0)=W(0)/\lambda$.
The time-evolution of the susceptibles is reconstructed based on the recovered individuals \eqref{for043}.
This requires to solve the same type of equation once again.
Relying on the balance equation \eqref{for037}, the result reads
\begin{align}\label{for001}
S(t)&=N(t)-I(t)-R(t)\,,
\end{align}
where similar to \eqref{for0052},
\begin{align}
\label{for002}
R(t)&=\left(N(0)-I(0)-S(0)\right)e^{-\rho t}+\gamma\int_0^te^{-\rho(t-r)}I(r)\,dr\,.
\end{align}
\end{subequations}
The initial values are approximated by steady state solutions of \eqref{for040}. That is, $I(0)=W(0)/\lambda$ and, provided mass-action \eqref{for004} applies, $S(0)=N(0)/\R_0$, where $\R_0=\beta(0)/\lambda$ defines the basic reproduction at $t=0$.
This number can be estimated by the ratio of ``births'' (expressed by the mean age at attack \cite{G74}, a) and ``deaths'' (mean life expectancy, $1/\mu$), $\R_0=1/a\big/\mu$.

{\em Second case:} $M(t)$ is given; $I(t)$ and $W(t)$ need to be calculated. Prevalence follows by $I=M/(q\lambda)$, incidence $W[I]$ is obtained from \eqref{for005} by differentiation. The reconstruction of $\beta(t)$ is now reduced to the first case.

The reconstruction is robust with respect to the initial values.
As \eqref{for0052} and \eqref{for002} reveal, the influence of the initial values vanishes exponentially with rates $\lambda$ and $\rho$, respectively.
That is, for large time ($t\gg1/\lambda$ and $\gg1/\rho$), $I$ and $R$ are determined exclusively by the convolution integrals.

In \eqref{for009}, contacts $\Sigma$ are not restricted to mass-action \eqref{for004}.
After specifying the functional $\Sigma[S,I]$, one may derive an expression for $\beta(t)$ that only depends on $W(t)$ or $M(t)$ and on an initial value (cf.~\cite{PWW11,H11,H12}).

\begin{figure}
\center\includegraphics[height=45mm]{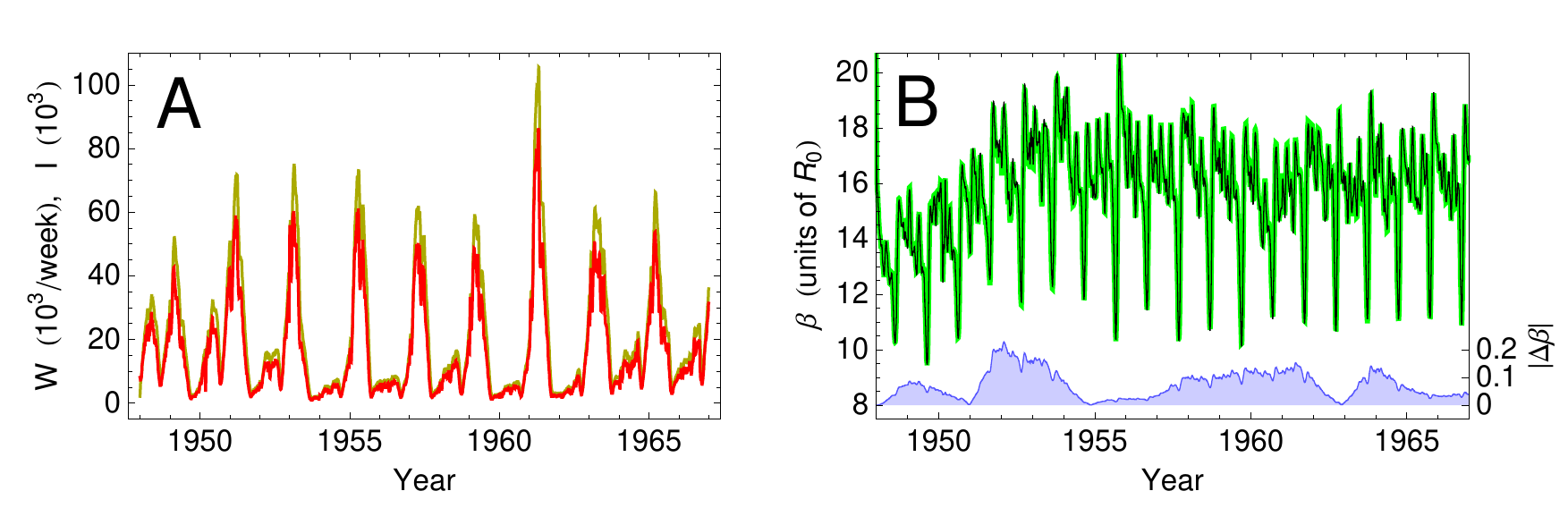}
\caption{\label{fig1} Data and transmission rates. The red curve in Panel A shows the weekly incidences of measles infections in England and Wales \cite{IIDDA}. Estimates of the corresponding disease prevalence (yellow) are computed according to \eqref{for0052}, assuming an infectious period of 9 days. In Panel B, the reconstruction of $\beta(t)$ is shown for constant (black) and time-dependent mortality (green). The constant mortality rate is estimated via the mean lifetime ($1/\mu=75\,\text{years}$, cf.~Fig.~\ref{fig2}D). Numerical differences between the reconstructed betas (blue) are very small, plotted in the lower part of the diagram. Time resolution has been increased from weeks to days.}
\end{figure}

\subsection{Including demographics}

So far we have assumed that the mortality rate $\mu$ and the removal rates, $\lambda$ and $\rho$, do not change over time.
The reconstruction formulas with time-dependent rates are very similar to the ones derived above,
only the exponents of \eqref{for0052} and \eqref{for002} are replaced by integrals
\cite{BD05}.
Applying our initial conditions, we obtain
\begin{subequations}\label{for00252b}
\begin{align}\label{for0052b}
I(t)&=\frac{W(0)}{\lambda(0)}e^{-\int_0^t\lambda(s)}ds+\int_0^te^{-\int_r^t\lambda(s)ds}W(r)\,dr\,,\\
\label{for002b}
R(t)&=\left(N(0)\left(1-\frac{1}{\R_0}\right)-\frac{W(0)}{\lambda(0)}\right)e^{-\int_0^t\rho(s)}ds+\gamma\int_0^te^{-\int_r^t\rho(s)ds}I(r)\,dr\,.
\end{align}
\end{subequations}
For measles, however, the improvement of the reconstruction resulting from these refined formulas is virtually invisible (Fig. \ref{fig1}B).

Unlike mortality, the birth rate $\nu$ ($=N'+\mu N$) and the population size $N$ have not been regarded as constant over time.
In contrast to the previous result, the impact of a changing population size
(discussed below) is clearly visible (Fig. \ref{fig2}C).

\subsection{Scaling invariance}

Provided contacts are modeled by mass-action \eqref{for004}, our setting is invariant with respect to {\em systematic errors} $\epsilon$ involved in the data. That is, 
the basic equations \eqref{for040} can be interpreted as a model for only a proportion of incidences $W_\text{eff}=\eta W$, susceptibles $S_\text{eff}=\eta S$, infectives $I_\text{eff}=\eta I$, and recovered individuals $R_\text{eff}=\eta R$ in an {\em effective population} $N_\text{eff}=\eta N$, where $\eta=1-\epsilon$ defines the {\em reporting proportion} with respect to the actual but unknown values.
This is easily confirmed multiplying each additive terms in \eqref{for040} by $\eta$.
    
The transmission rate obtained for such an effective description is the same as for the actual values, $\beta_\text{eff}=\beta$. This can be seen by expanding the basic equation \eqref{for009},
\begin{align}\label{for006}
\beta=\frac{N~W}{S~I}=\frac{\eta N~\eta W}{\eta S~\eta I}=\beta_\text{eff}\,;
\end{align}
the additional ''$\eta$'' cancel.

The reporting proportion is obtained from \eqref{for041} --- by substitution of \eqref{for038} and \eqref{for009},
$S'=\nu-\mu S+\delta R+\Theta-W$,
by subsequent integration over time (up to some large $t$, so that the left-hand side becomes small in comparison with the terms on the right-hand side), $0\approx\int_0^t S'(r)\,dr=\int_0^t\left(\nu(r)-\cdots-W(r)\right)dr$,
and by some infection-specific simplifications.
For measles (i.e., $\delta\approx0$, $\Theta\approx0$, $\R_0\gg1$),
the integral simplifies to
$0\approx\int_0^t\left(\nu(r)-W_\text{eff}(r)/\eta\right)dr$
so that the ratio between cumulated incidences and births (cf.~\cite{FG00}),
% or a time-average of
\begin{subequations}
\begin{align}\label{for047}
\eta_\nu(t)=\frac{\int_0^tW_\text{eff}(r)\,dr}{\int_0^t\nu(r)\,dr}\,,
\end{align}
as well as moving averages $\overline\eta_\nu(t)$ of this ratio, yield estimates of the reporting proportion $\eta$.
In numerical simulations we utilize the harmonic mean to calculate $\overline\eta$. If, for example, the average is taken over the whole time period then $\overline\eta_\nu=n\big/\sum_{i\le n}\nu(t_i)/W_\text{eff}(t_i)$, where $n$ is the number of data points $t_i$.

\begin{figure}
\center\includegraphics[height=130mm]{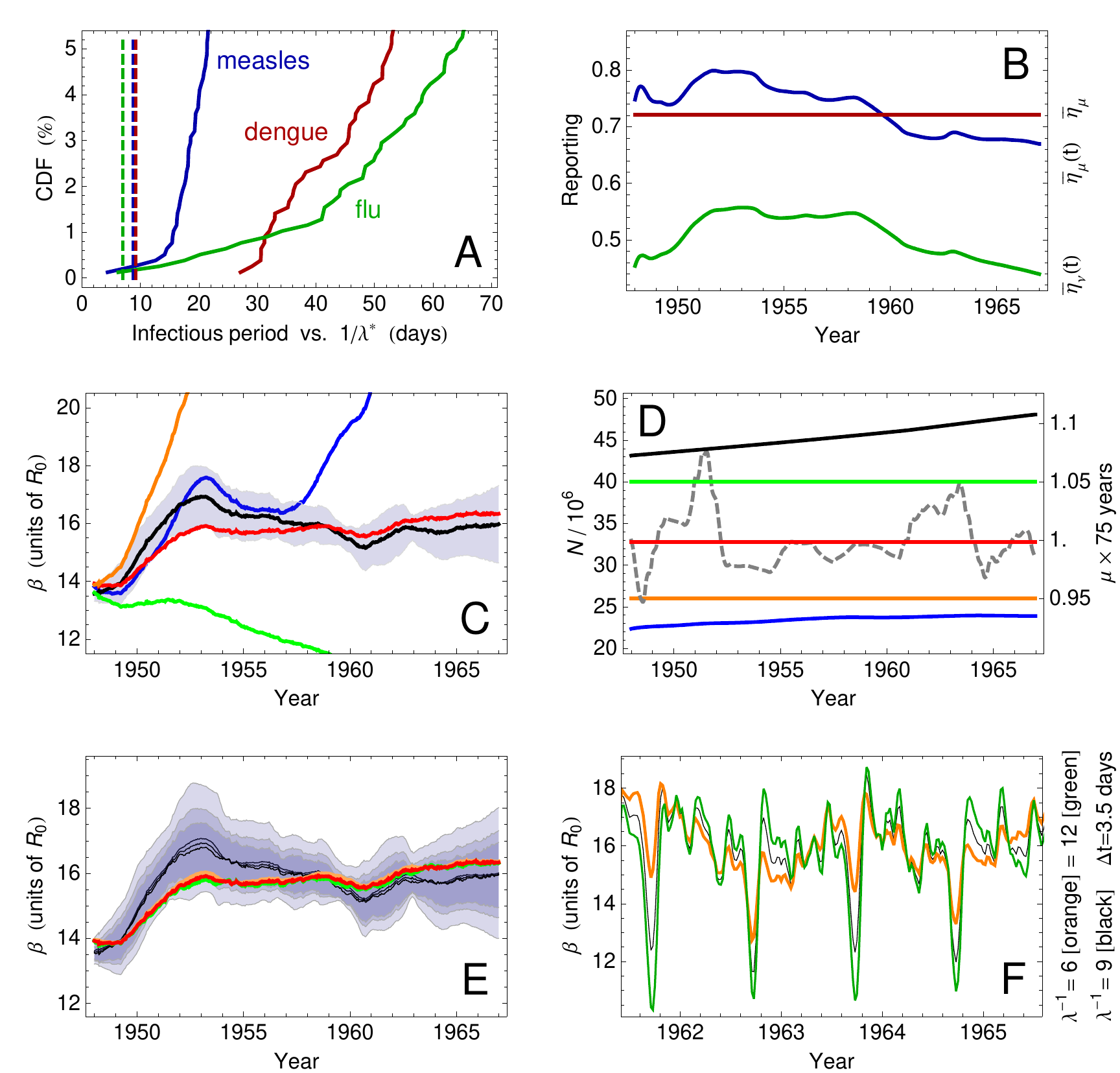}
\caption{\label{fig2}Parameters. For each of the three infections studied in the article, Panel A shows the cumulative distribution function (CDF) of the lower bound $\lambda^\ast$ of the removal rate \eqref{for012} with respect to the data points. The mean infectious periods used in our numerical simulations are indicated by dashed lines. The Panels B--F refer to measles. In B, the two kinds of reporting proportions --- based on natality (green) and on mortality (blue) --- are plotted over time. The red line indicates the reporting proportion that corresponds to a constant effective population (of size $32.8\times10^6$).
In Panel C, the reconstruction results (2-year moving averages) are shown for the population sizes plotted in Panel D: the actual size (black), a slowly varying size (blue), and three constant sizes (green, red, orange).
In respective order, the reconstructions are performed based on reporting proportions $\overline\eta_\nu(t)$, $\overline\eta_\mu(t)$, and $\overline\eta_\mu$; the mortality rate (dashed) is plotted in Panel D too. In the two remaining panels, E and F, the mean infectious period is varied: 6~days (orange, light gray), 9~days (red, gray), 12~days (green, dark gray). Panel E shows reconstruction results (2-year moving averages) for a constant (colored) and a variable population size (black; reporting errors are indicated gray). Panel F shows reconstruction results (1-month moving averages) for a variable population size; amplitudes are higher and more regular for longer infectious periods.}
\end{figure}

For measles, the scaling behavior is of particular importance.
For the pre-vaccination data from England and Wales the reporting proportion is
known to be significantly smaller than one \cite{FC82,FG00}.
If we therefore apply the reconstruction \eqref{for00252a} to raw data, the method does not work.

\subsection{An effective population of constant size}

Whenever our knowledge about the demography is insufficient, we propose that the size of the population is constant over time, $N(t)=\overline{N}$. Even for simplicity we could make such an assumption. However, it is not trivial to estimate an appropriate constant effective size. Scaling invariance together with \eqref{for047} does not work. For our measles example, the resulting size of $\eta_\nu\overline N\approx0.5\times46\times10^6=23\times10^6$ is far too low.

If the population size is chosen too low, the reconstruction produces unrealistically high beta values and eventually collapses (as happening with $N_\text{eff}=23\times10^6$, cf. orange curve in Fig. \ref{fig2}C).
This is not surprising when looking at the reconstructed susceptibles $S_\text{eff}=N_\text{eff}-I_\text{eff}-R_\text{eff}$ in the denominator of \eqref{for006}. By lowering $N_\text{eff}$ the number of susceptibles $S_\text{eff}$ approaches zero and eventually turns negative.
If, on the other hand, the population size is chosen too large, the reconstructed beta decreases below reasonable values (cf. green curve in Fig. \ref{fig2}C) and, as \eqref{for006} reveals, tends towards the trivial solution, $\beta=W_\text{eff}/I_\text{eff}\approx\lambda$.
The sketched behavior is quite sensitive and can therefore be used to narrow down an appropriate size for the effective population.
Assuming that $\beta(t)\approx\text{const}$, we obtain $N_\text{eff}\approx33\times10^6$. This number is much smaller than the number of inhabitants ($46\times10^6$) in England and Wales at the time \cite{HA99}.

When proposing a constant population size a straightforward application of \eqref{for047} is problematic.
{\em First}, Equation \eqref{for047} contains the birth rate $\nu$ but the reconstruction equations \eqref{for00252a} rely on the death rate $\mu$. {\em Second}, A constant population requires that $N'=0$ (i.e., $\nu=\mu N$), even if in reality the size changes.
These inconsistencies are solved, however, when replacing $\nu$ in \eqref{for047} by $\mu N$.
The effective population is then given by
\begin{align}\label{for007}
N_\text{eff}(t)=\frac{\int_0^tW_\text{eff}(r)\,dr}{\int_0^t\mu(r)\,dr}\,.
\end{align}
\end{subequations}
Here, $\eta_\mu={N}_\text{eff}/\overline{N}$ ($\neq\eta_\nu$) defines the reporting proportion, but again one must employ suitable averages.
For our measles data we obtain ${N}_\text{eff}=32.8\times10^6$
when proposing a constant effective size during the period 1948--67; see the reconstruction results (red curves) in Figures \ref{fig2}C and \ref{fig3}.

Alternatively, one might start with a moving average $\overline{N}_\text{eff}(t)$ of \eqref{for007} and apply the reconstruction to raw data $W(t)$. Such a procedure works, although it does not seem to be robust (blue curve in Fig. \ref{fig2}C).

\subsection{Numerical simulations}

When modeling real data, the functions $W(t)$ and $M(t)$ are given at a certain resolution in time (e.g., $t_{n+1}-t_n=1\,\text{week}$).
Therefore, it must be explained how continuous-time formulas are applied to functions defined an a discrete time line $\{t_n\}_{n\in\mathbb{N}}$.
The derivative of a function $F(t)$ at $t=t_n$ is approximated by $F'(t_{n})=\frac{F(t_{n+1})-F(t_{n-1})}{2\Delta t}$.
This definition works well as long as infectious periods are short and removal rates $\lambda$ are large.
Admissible values are given by
\begin{align}\label{for012}
\lambda\ge\lambda^\ast=\sup_{n}\frac{|I'(t_n)|}{I(t_n)}\,,
\end{align}
imposing that incidence rates are non-negative; see \eqref{for005}.
The values ($1/\lambda=9,9,7~\text{days}$) utilized for the infections in this article (measles, dengue, flu) fulfill this condition for nearly all data points ($>99\%$), and $\lambda\gg\lambda^\ast$ for most of the data points (Fig. \ref{fig2}A).

Numerical simulations indicate that the specific choice of $\lambda$ is not important on a large time scale (cf. the 2-year moving average of beta in Fig. \ref{fig2}E). This result, which exempts us from applying other methods to determine precise infectious periods, is confirmed by the following approximate equation and its $\lambda$-independent right hand side,
\begin{align}\label{for049}
\frac{\beta(t)}{\lambda}\approx\frac{N(t)}{N(t)-N(0)\left(1-\frac{1}{\R_0}\right)e^{-\rho t}-\int_0^te^{-\rho(r-t)}W(r)dr}\,.
\end{align}
The equation is derived from \eqref{for006} and \eqref{for002} by incorporating the estimates:
$W\approx\lambda I$, $\lambda\approx\gamma$, $I\ll N$.

The reconstruction formulas, \eqref{for0052} and \eqref{for002}, contain integrands of the form $e^{\kappa r}F(r)$. Their discrete versions require particular care,
especially when time resolution is small (i.e., $\kappa\Delta t\gg0$) and the exponential term dominates. We utilize the Gaussian quadrature rule,
\begin{align}\label{for044}
 \int_0^te^{\kappa r}F(r)\,dr=\sum_n w_n e^{\kappa t_n}F(t_n)\,,
\end{align}
with an index independent weight $w_n=\frac{1-e^{-\kappa\Delta t}}{\kappa}<\Delta t$. The weight accounts for the exponential contribution in between observations, imposing that $\int_{t_n-\Delta t}^{t_n}e^{\kappa r}F(t_n)dr=w_n e^{\kappa t_n}F(t_n),~\forall n$.

Usually integrals are approximated well by summation \eqref{for044} with the trivial weight, $w_n=1$. Here this would lead to huge errors:
the integral in \eqref{for0052}, for example, must be corrected by $w_n\approx2/3$, if employing weekly data and an infectious period of one week (where $\kappa\Delta t\approx1$).
This might be a reason why the method has not been proposed in the literature
before.
Another similar reason might be the non-trivial scaling behavior discussed in the two previous subsections.
In \cite{PWW11,M13}, the scaling problem has not been addressed. 
By computing ``$\beta/N$'' the authors avoid the problem of adjusting $N$; they only need to estimate the initial value $\beta(0)/N(0)$.
While being apparent in numerical simulations, scaling is absent in theoretical calculations and therefore unnoticed in \cite{H11,H12}.

\subsection{Computations}

The eight figures in this article can be re-computed in less than 20\,min.
On a usual laptop, our Mathematica \cite{math6} implementation

{\small
\begin{verbatim}
                                                                                      (* INPUT *)
   wL=measles_data;                            (* a list of weekly incidences, e.g., from [23] *)
   R0=15; n=32.8*10^6;                                (* initial R0, effective population size *)
   lambda=7/9.; rho=1/75/52.; gamma=lambda-rho;         (* weekly removal, mortality, recovery *) 
                                                                      (* SUBROUTINES/FUNCTIONS *)
   weight[rate_]:=(1-Exp[-rate])/rate;                                       (* rate in 1/week *)
   expL[rate_]:=Table[Exp[-rate*t],{t,T}];                                     (* exp(-rate*t) *)
                                                                        (* BETA RECONSTRUCTION *)
   T=Length[wL];                                                           (* # of data points *)
   iL[0]=expL[lambda]*wL[[1]]+weight[lambda]*                            (* infectives, Eq. 9a *) 
         Table[Sum[Exp[lambda*(r-t)]*wL[[r]],{r,t}],{t,T}]; 
   iL[1]=wL-lambda*iL[0];                             (* time-derivative of infectives, Eq. 8  *) 
   rL=expL[rho]*((1-1/R0)*n-iL[0][[1]])+gamma*                            (* recovered, Eq. 9c *)
      weight[rho]*Table[Sum[Exp[rho*(r-t)]*iL[0][[r]],{r,t}],{t,T}];
   sL=n-iL[0]-rL;                                                      (* susceptibles, Eq. 9b *)  
   betaL=n*wL/sL/iL[0];                                                      (* OUTPUT, Eq. 7  *)
\end{verbatim}}

\noindent
takes about 2\,sec to reconstruct beta for 1,000 data points (Figs.~\ref{fig3}A, \ref{fig4}A) and about 100\,sec for 7,000 points (Figs.~\ref{fig3}C, \ref{fig4}B).
Because of the convolution integrals in \eqref{for00252a},
the computation time increases with the square of the data size.

\subsection{Error estimates}

\begin{figure}
\center\includegraphics[height=83mm]{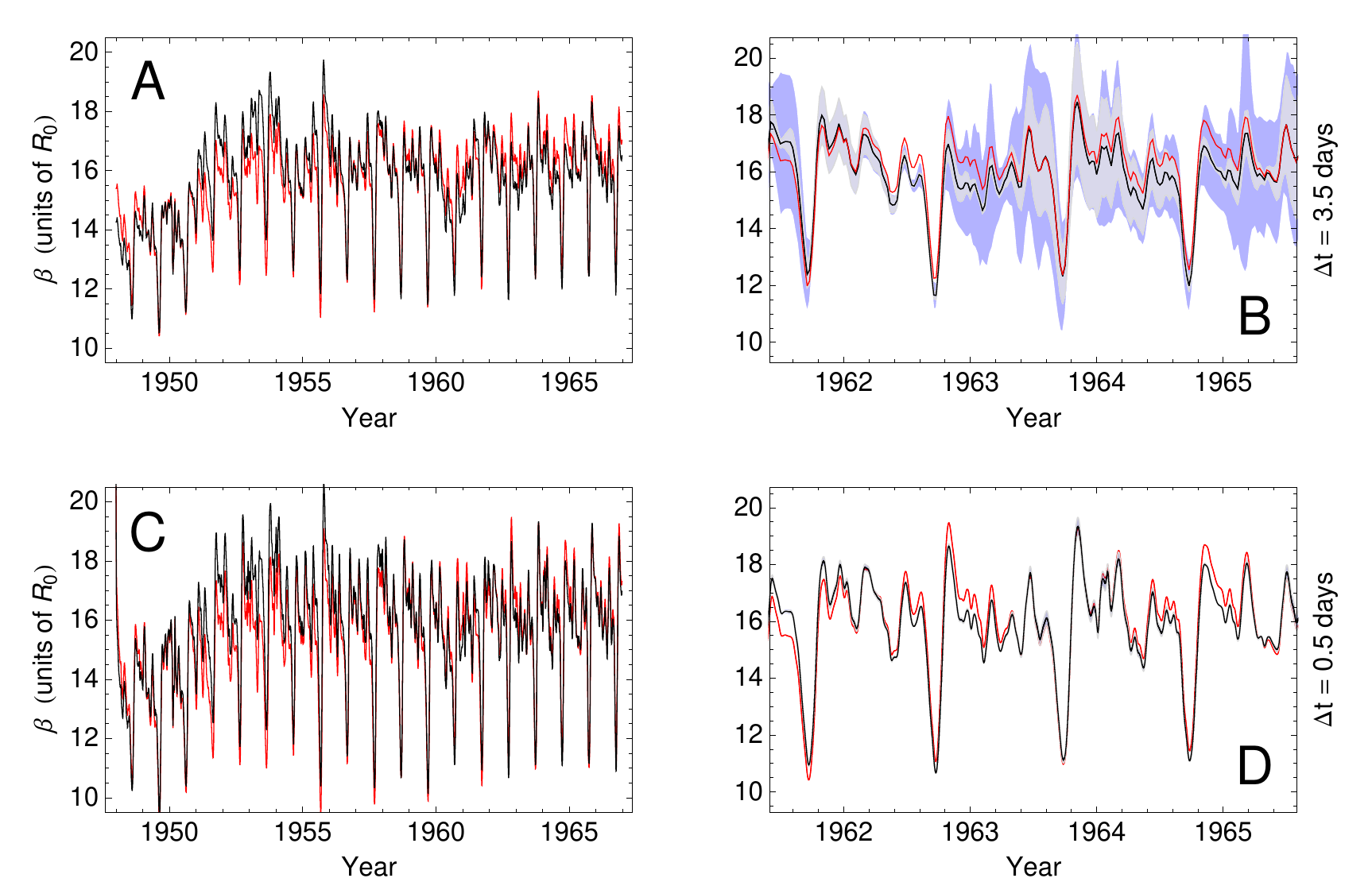}
\caption{\label{fig3} Errors. Transmission rates are reconstructed for a variable (black) and a constant population size (red, $N_\text{eff}=32.8\times10^6$).
The resulting curves are very similar. For the original (1-week) resolution, shown in Panels A and B, the deviation stays within the error bounds.
Errors due to reporting (gray) and time resolution (blue) --- both estimated for a variable population size --- are shown in the panels on the right hand side. 
Panels C and D show reconstructions for an increased resolution (days); interpolation is done via cubic polynomials. Here the deviation exceeds the tiny error bounds (in Panel D, barely visible). The reconstruction at high resolution produces more uniform looking curves, $\beta(t)$, with higher amplitudes than the reconstruction at the original resolution. The reconstruction at high resolution has an annual instead of a biennial period. 
}
\end{figure}

For our measles data, the transmission rate $\beta(t)$ is known to have an annual period \cite{FC82}.
The reconstruction, however, is showing frequency modes with periods longer than one year, especially when the time resolution is low (cf.~Figs.~\ref{fig3}B, \ref{fig4}AC).
Interpolation of data successfully removes the extra modes, suggesting that these modes are caused by an insufficient data resolution.

Another major source of error is the reporting proportion, which we calculate by a version of \eqref{for047}.
For both sources of error we determine the error of beta after the reconstruction (Fig.~\ref{fig3}).
For simplicity we only consider the case where $I$ ($\approx W/\lambda$) is given and $R$ is computed according to \eqref{for002}.
We assume that $\eta_\nu(t)\approx\overline\eta_\nu$ and $|I'(t)|/I(t)\ll\lambda$.
Taylor expansion of \eqref{for009} then yields
\begin{subequations}\label{for0100}
\begin{align}\label{for010}
|\Delta\beta(t)|=~&
\frac{\left|1-R(0)e^{-\rho t}+R(t)\right|}{\lambda\,\overline\eta_\nu N(t)}\,
\beta(t)^2\,|\Delta \eta_\nu(t)|
+\bigO(|\Delta\eta_\nu(t)|^2)\\\label{for023}
&+\left(1+\frac{\gamma}{\rho}\,(1-e^{-\rho t})\right)
\left|2-\frac{\lambda}{\beta(t)}-\frac{2\,R(0)}{e^{\rho t}N(t)}\right|
\frac{\overline\eta_\nu\beta(t)^3}{\lambda^2N(t)}\,|\Delta\eta_\nu(t)|\,|\Delta I(t)|\\\label{for024}
&+\left(1+\frac{\gamma}{\rho}\,(1-e^{-\rho t})\right)\frac{\overline\eta_\nu\beta(t)^2}{\lambda\,N(t)}\,|\Delta I(t)|+\bigO(|\Delta I(t)|^2)
\end{align}
\end{subequations}
as $|\Delta\eta_\nu(t)|\to0$ and $|\Delta I(t)|\to0$. 
The error of the reporting proportion is estimated by the deviation from the mean, $|\Delta\eta_\nu|=|\eta_\nu-\overline\eta_\nu|$.
The error of the data resolution is estimated by the error related to the time resolution, $|\Delta I|=|I'|\,\Delta t+\bigO((\Delta t)^2)$, where $\Delta t_{n}=(t_{n+1}-t_{n})/2$.

\section{Results}

For three sets of infectious disease data we reconstruct beta as a function of time. Each set represents an epidemiological scenario for which the method can provide new insights into the transmission dynamics.
Epidemiological results are discussed immediately, methodological aspects are reviewed in the Discussion.

\subsection{Measles}

\begin{figure}
\center\includegraphics[height=105ex]{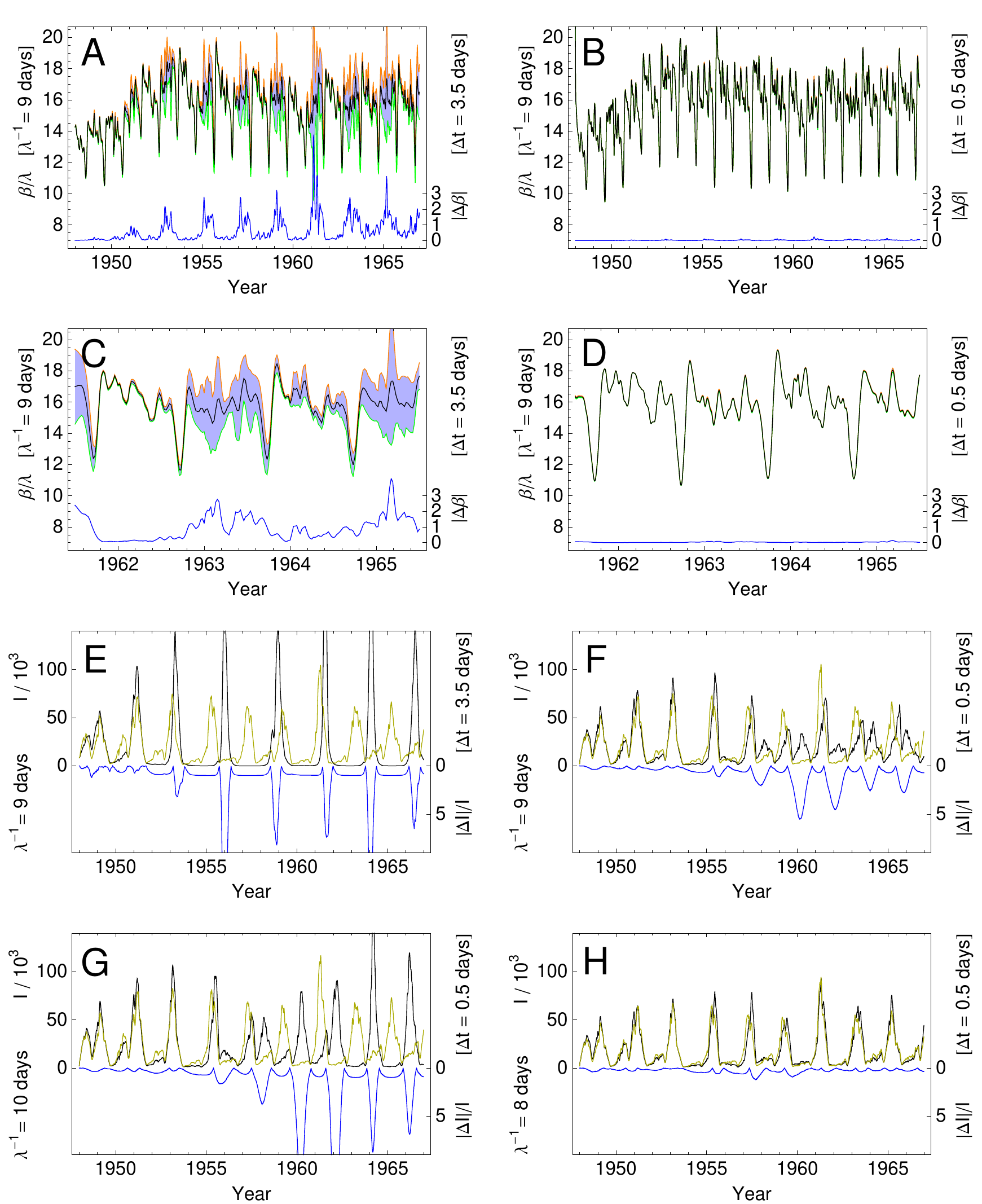}
\caption{\label{fig4} Beta reconstruction and data reproduction.
Panels A,C show $\beta(t)$ reconstructed from weekly incidences of measles in England and Wales (cf.~Fig.~\ref{fig1}A).
The infectious period is assumed to be 9 days.
Panels B,D illustrate the reconstruction for a 1-day resolution, obtained by interpolation of the original data via cubic polynomials.
Panels C,D zoom into Panels A,B so that the three school terms of high transmission and the drop in summer can be seen.
The peak shortly after the Christmas holidays has been smoothed out by the plotting resolution of 1 month.
At the original (1-week) resolution, the biennial period --- induced by the pattern of incidence --- is visible.
For a 1-day resolution, this is different: the beta curves of odd and even years are similar.
Errors due to insufficient time resolution are indicated blue; these errors vanish at high resolution.
In Panels E-H, prevalence data $I(t)$ are reproduced from the reconstructed  $\beta(t)$. Reproduction results profit from higher resolution (Panels E,F) and from shorter infectious periods (Panels G,H). For a 1-day resolution and an 8-day infectious period (Panel H), the data are reproduced almost identically.
}
\end{figure}

Transmission rate reconstructions have taught us about the epidemiology of measles. The first result  \cite{FC82} was obtained with data from the pre-vaccination era in England and Wales (Fig.~\ref{fig1}A). Biennial patterns of incidence were related to transmission events with an annual period. In particular, the three school terms and the Christmas holidays
were identified as periods of high transmission (cf. our reconstructions in Fig.~\ref{fig4}CD).
Subsequent work \cite{BFG02,MF05,CF08} introduced probabilistic methods --- to achieve better parameter estimates and, being able to control errors, to put the beta reconstruction on a more trustworthy foundation.
The more recent work \cite{HERE11,WCBIL12,WYCIL13} combines nonlinear optimization techniques and compartmental models, which can significantly shorten the computation time \cite{WCBIL12}.

Our approach is similar to the classical one \cite{FC82}.
In that it is based on SIR dynamics (\ref{for040}) and requires the reconstruction of the involved compartments. In contrast to the discrete-time approach of \cite{FC82}, we use continuous-time expressions and work with differential equations, similar to \cite{PWW11,H11}.
For our numerical simulations we define a discrete calculus, including a refined representation \eqref{for044} of the convolution integrals, which model the time evolution. Initial conditions of \eqref{for0052} and \eqref{for002}
are chosen in accordance with steady state solutions, where \eqref{for002} employs the initial value of the basic reproduction.
In practice, we choose the initial value 
to match the basic reproduction at the end of considered time period (end of 1966), where the mean age at attack and the mean life expectancy are available most accurately \cite{G74}.

The reconstruction requires knowledge of the mortality distribution, although knowing the life expectancy turns out to be sufficient (Fig.~\ref{fig1}B).
The mean infectious period \eqref{for012} is another input parameter.
The reconstruction, however, does not rely much on its precision (Fig.~\ref{fig2}EF).
If demographic information is available, as being the case for England and Wales \cite{HA99}, the distribution of births is used to estimate the reporting proportion \eqref{for047}. This parameter is needed for calibration; it represents the epidemiological data --- which usually suffer from systematic errors --- as effective numbers \eqref{for006}.
Otherwise, if demographic information is incomplete,
one assumes that the population is of constant size, determined by the mean life expectancy \eqref{for007} or, even simpler, by the requirement of convergence. The size of the effective population happens to be very sensitive (Fig.~\ref{fig2}C), allowing for such a criterion. 
Both our reconstructed betas --- based on a variable (black) and on a constant population size (red) --- resemble the curves known from the literature. The deviation between our two curves is small (Fig.~\ref{fig3}) and stays within the error bounds of the original resolution (Fig.~\ref{fig3}B).

The periodic behavior of $\beta(t)$ tells us about another parameter, the infectious period.
We find that for short infectious periods ($1/\lambda<10\,\text{days}$) the reconstructed transmission rates show biennial frequency modes (Fig.~\ref{fig4}A): in 1962 and 1964 the beta curves are similar (beta decreases on average), in 1963 and 1965 the curves are similar too (but beta increases), whereas in 1962 and 1963 the curves look different (Figs.~\ref{fig3}B,~\ref{fig4}C).
Even if prevalence of measles infections is known to show various frequency modes \cite{ERB00}, modes of the transmission rate that are larger than one year are unrealistic (at least for the U.K. \cite{GGP99}). 
Therefore, the result must represent an artifact, which we assume is caused by the poor resolution of the original data. In fact, errors are large every second year (Figs.~\ref{fig3}B,~\ref{fig4}C). If, by interpolation, we increase the resolution from weeks to days, we observe that biennial modes change into annual modes (cf.~Figs.~\ref{fig3}D,~\ref{fig4}D).
The same effect, however, can be achieved by employing the original data with longer yet more unrealistic infectious periods.

How about the ability to reproduce the original data?
We perform the reproduction by solving the basic equations
(\ref{for041},\,\ref{for042})
employing the reconstructed betas together with the parameters used for the reconstruction (Fig.~\ref{fig4}E-H).
Here, shorter infectious periods turn out to be in favor (Fig.~\ref{fig4}G vs.~\ref{fig4}H); the quality of the reproduction in Figure \ref{fig4}H is striking.
Similar to the reconstruction, reproduction results are satisfactory only at high resolution (Fig.~\ref{fig4}E vs.~\ref{fig4}F).
The poor time resolution in \cite{FC82} might thus explain why the reproduction of data (from betas computed in \cite{FC82}), investigated in \cite{MD93}, did not work so well.
The same problem might occur when reproducing data based on
betas obtained in \cite{PWW11}; their beta curves seem to have a low resolution as well.

We conclude that there is a trade-off between the quality of the reconstruction and the quality of the reproduction balanced by the length of the infectious period.
Working with higher resolution --- obtained by interpolation of the original data --- is beneficial for both, reconstruction and reproduction.
This is not surprising, as the assumption underlying the applicability of interpolation is smoothness, being consistent with our methodology relying on differential equations.
Good estimates of the infectious period, however, are relevant only for the short-term behavior of beta (Fig.~\ref{fig2}F). Long-term, the length of the infectious period is not important (Fig.~\ref{fig2}E).

\begin{figure}
\center\includegraphics[height=30ex]{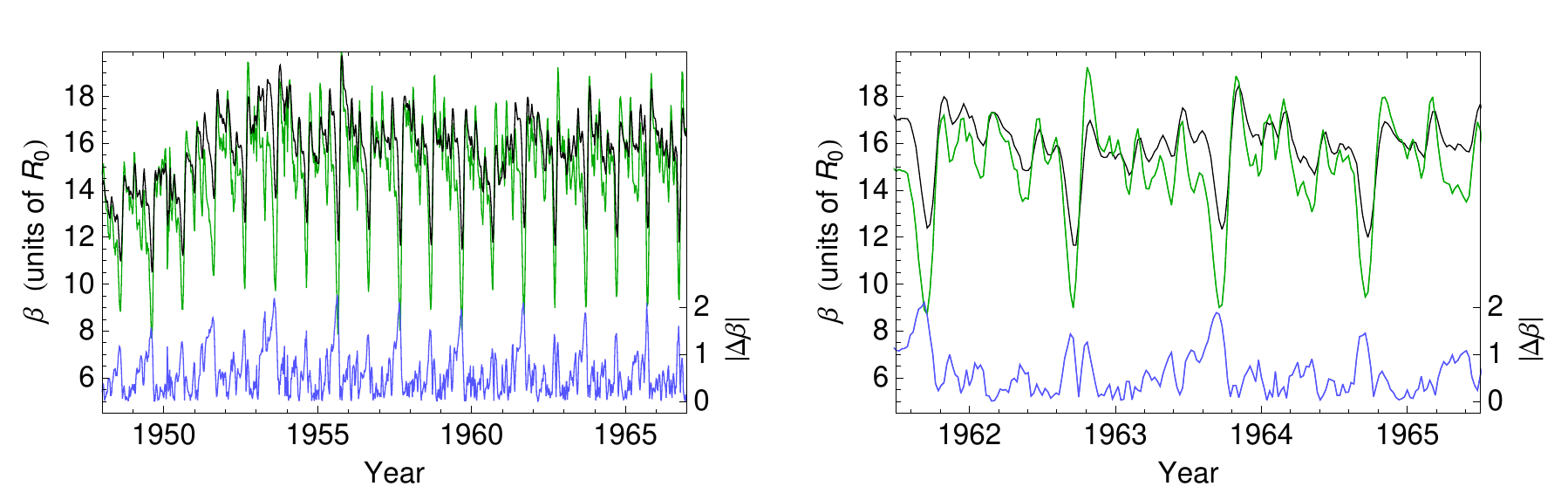}
\caption{\label{fig5} SIR versus SEIR reconstruction. The diagrams show $\beta(t)$ reconstructed for SIR (black) and SEIR systems (green) with mean incubation period of $1/\varepsilon=5\text{ days}$. 
The deviation between the beta values is plotted (blue) in the lower part of the diagrams. For both models, the mean infectious period is 9 days. Computed at original (1-week) resolution, the SIR reconstruction shows an artifactual two year period, whereas the SEIR reconstruction does not.}
\end{figure}

To widen the applicability of the method we start to extend the reconstruction to models with multiple infective stages.
For measles, it is common to also introduce a compartment of exposed (infected and not yet infective) individuals $E$ with removal rate $\varepsilon$. Accordingly, we replace Equation \eqref{for042} by
\begin{subequations}
\begin{align}
E'&=\beta SI/N-(\varepsilon+\mu)E\\
I'&=\varepsilon E-\lambda I\,;
\end{align}
\end{subequations}
incidence is then given by $W=\varepsilon E$.
In doing so we reconstruct beta according to an SEIR model, which relies on more realistic assumptions than an SIR model.
Consequently, the result (Fig.~\ref{fig5}) is more precise, it resembles the beta curve of the SIR-model (Fig.~\ref{fig3}CD) reconstructed at higher time resolution.

\subsection{Dengue}

Another infection that involves multiple infective stages is dengue. This is a vector-born infection caused by an RNA virus that occurs in four serotypes with oscillating prevalence in the human host-population. Primary infections are acute and only last for a couple of weeks. Individuals are protected against reinfection with the same serotype strain but live at risk to become severely ill and develop symptoms referred to as dengue hemorrhagic fever (DHF) when reinfected with another serotype. This phenomenon is known as an antibody-dependent enhancement (ADE) effect \cite{FAG99}.

Recently, dengue has been reported to re-emerge in some regions of the world, although transmission rates have successfully been reduced by vector-control measures  \cite{TNS08}.
The apparent contradiction has been explained by two competing mechanisms: clinical cross-protection \cite{NK08} and demographic transitions \cite{CIL08}. 
The first approach focuses on host-pathogen interactions, the second one on changes in the host population.
But which of two offers the correct explanation? This is the question we try to answer in this section utilizing the beta reconstruction.
Different from both these approaches, our reconstruction relates DHF data and transmission rates via a formal procedure.
This may falsify one of the suggested explanations.

Three compartments in the human host-population are relevant for the dengue epidemic: susceptible, infected, and exposed individuals.
A model based on the contact functional $\Sigma[S,I]$ is given by
\begin{subequations}\label{for032}
\begin{align}
S'&=N'+\mu(N-S)-\beta \Sigma[S,I]\label{for032a}\\
I'&=\beta\Sigma[S,I]+\beta k\Sigma[E,I]-\lambda I\label{for032b}\\
E'&=F[I]-\mu E-\beta k\Sigma[E,I]\,,\label{for032d}
\end{align}
where strain-specific information is encoded by the number $k$.
The functional $F$ represents the replenishment of exposed individuals.
The {\em exposed}\, compartment refers to individuals who were infected at least once and have therefore been exposed to one or more serotypes. These individuals are also susceptibles, even if becoming reinfected with a new serotype strain is  less likely than getting infected for the first time;
the parameter $k~(<1)$ models the likelihood of reinfection.
We do not need to estimate $k$ via probabilistic arguments or extra data, the implementation is achieved through consistency requirements explained below.

In \cite{NK08}, removal of infected individuals into the exposed compartment has been proposed to be time-delayed and reduced by cross-protection. This mechanism can be modeled by two additional compartments,
\begin{align}\label{for032e}
R'&=(\lambda-\mu)I-(\rho+\mu) R-\beta l\Sigma[R,I]\\
P'&=\beta l\Sigma[R,I]-\mu P\,,
\end{align}
where $R$ represents temporarily and $P$ permanently immunized individuals.
The contact term $\beta l\Sigma$, which corrects for multiple serotypes, accounts for events of permanent immunization,
the term
\begin{align}\label{for015}
F[R]=\rho R
\end{align}
\end{subequations}
quantifies the rate $F[R[I]]$ at which the exposed compartment is replenished.

\begin{figure}
\center\includegraphics[height=72ex]{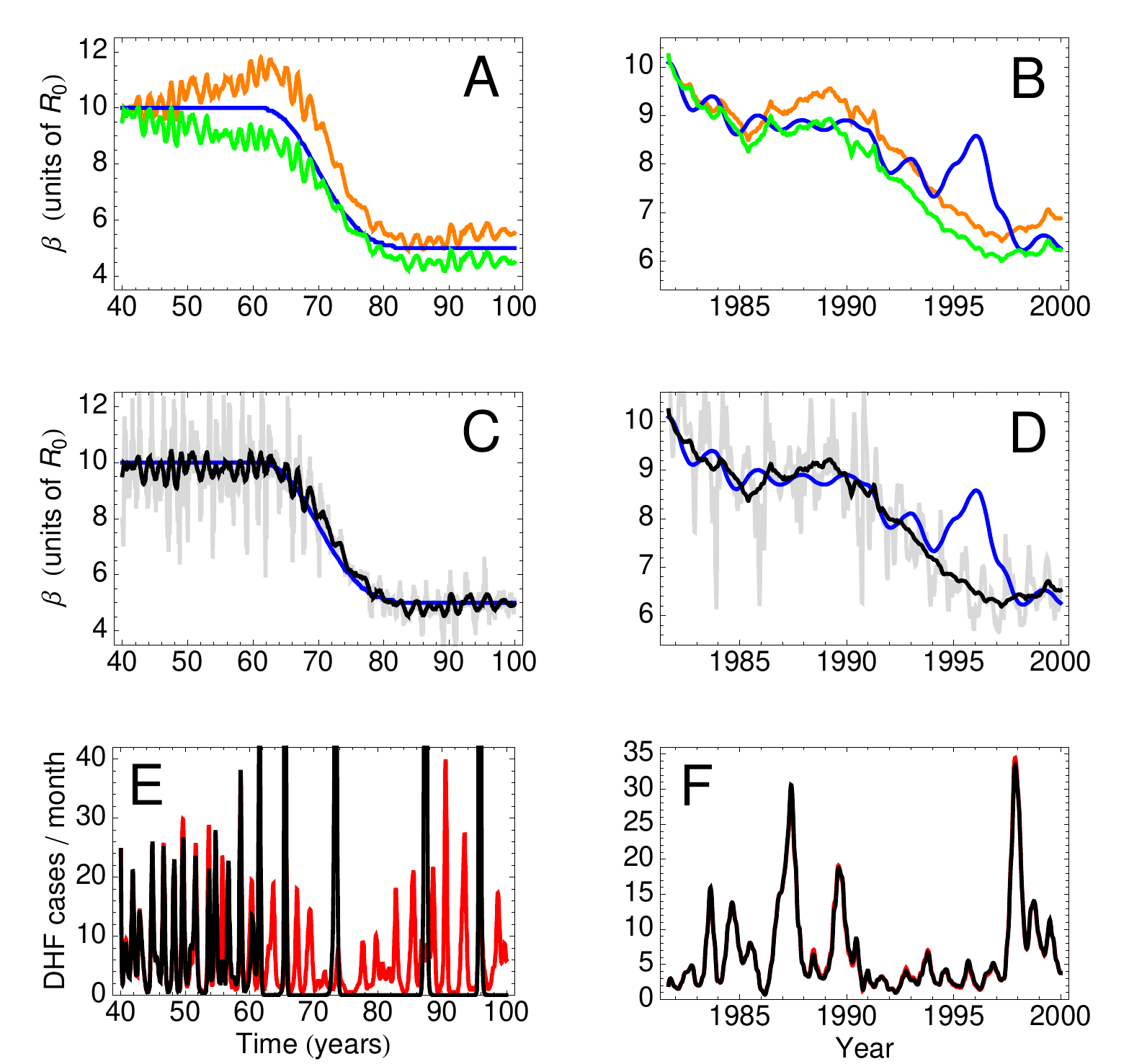}
\caption{\label{fig6}Dengue. The diagrams on the right hand side represent data from Thailand, the ones on the left hand side a model scenario. Red curves show DHF data, blue curves the transmission rates. These curves are taken from Figure 4 in \cite{NK08} (at a resolution of $220\times160$ and $700\times160$ sq.~pixels, for real and model data, respectively).
Panels A and B show the reconstruction results for scaling factors $q$ deviating from the optimal values ($q/p_2=2.55$ and $2.19$, resp.) by $\pm5\%$ (orange/green).
In accordance with \cite{NK08}, the risk of developing DHF is assumed to be $p_2=.0338$, and the mean infectious period $1/\lambda=9\,\text{days}$.
The black curves in Panels C and D are reconstructed from DHF data (red curves in Panels E and F).
The result is nearly perfect, except for the beta value in 1996 (Panel D). The blue curve in D is computed based on the mean age of the infected individuals suffering from DHF \cite{NK08} (i.e., the blue curve is not related to the red curve in Panel F by computation). The reconstructed betas are plotted gray, their half-year moving average black. DHF cases (black curves in Panels E and F) are reproduced by employing the reconstructed betas following the exposed SIR-dynamics (explained in the text). The data curve (Panel F) is almost identically reproduced (red spots are barely recognizable). For the model scenario (Panel E), reproduction is good for the first 10--20 years. }
\end{figure}

Similar to the parameter $k$, we do not need to estimate $l$ or $\rho$ to reconstruct beta. We only estimate the likelihood that relates DHF rates $M$ and incidences of secondary infection,
\begin{align}\label{for033}
q=\frac{M}{\lambda H}\,,
\end{align}
where $H=hI$ (with $h<1$) represents the infected individuals with prior exposure.
It is sufficient to consider secondary infections here, as DHF data rarely account for cases resulting from primary infection: $p_1\lambda (I-H)+p_2\lambda H\approx p_2 \lambda H$, as
$h\gg p_1/p_2$. %=.059\,,
Due to ADE, the risk of developing DHF after re-infection is
$p_2/p_1=16.9$
times higher than without prior exposure; see legend to Figure 3 in \cite{NK08} and Table \ref{tab1}.

The unknown DHF-likelihood $q$ is determined by the global behavior of $\beta(t)$.
For dengue in Thailand, beta is known to decline (blue curve in Fig. \ref{fig6}D), corresponding to an increasing mean age of the individuals suffering from DHF \cite{NK08}.

It may surprise that $q$ is all that is needed to reconstruct the correct transmission rates; our beta curve (black) successfully fits the (blue) reference curve in Figure \ref{fig6}CD.
We computed $\beta(t)$ following the second case in Methods while using mass-action \eqref{for004}.
The two sets of dengue data (in Fig. \ref{fig6}) are taken from \cite{NK08}, representing DHF cases in Thailand (right) and a simulated model scenario (left).

To understand how the method works one must realize that $\beta(t)$ is not reconstructed literally --- via $S$ and $E$, as defined in \eqref{for032}. The method keeps track of the exposed individuals $E$ and reconstructs a truncated version of \eqref{for032b},
\begin{align}\label{for045}
H'&=\beta EK/N-\lambda H\,,
\end{align}
where $K=kI$ models the compartment of individuals that infect exposed individuals.
When expressing replenishment of exposed individuals \eqref{for015} by a multiple of natality,
\begin{align}\label{for046}
F[N]=\eta\,\mu N
\end{align}
(given by mortality, as the reconstruction is done for a constant effective population),
one obtains the relevant subset of \eqref{for032},
\begin{subequations}\label{for032EkI}
\begin{align}
\frac{E'}{\eta}&=\mu N-\mu\frac{E}{\eta}-\beta\,\frac{E}{\eta}\frac{K}{N}\\
K'&=\beta\,\frac{E}{\eta}\frac{K}{N}-\lambda K\,.
\end{align}\end{subequations}
Comparison with the basic SIR equations (\ref{for041},\,\ref{for042}) reveals the following correspondences, $E/\eta\to S$ and $K\to I$, where 
\begin{align}
k=\frac{h}{\eta}\quad\text{and}\quad\eta=\frac{q}{p_2}\,.
\end{align}
These definitions demonstrate that the DHF-likelihood $q$ is the only parameter required as input.
Numerical simulations (Fig. \ref{fig6}AB) determine its value at 2--3 times the risk of developing DHF after re-infection (cf. Table \ref{tab1}).

\begin{table}\centering
{\small
\begin{tabular}{|c||c|c|}\hline
& model data & real data\\ \hline\hline
$\eta$&2.19&2.55\\ \hline
$1/\rho$&52&25\\ \hline
$\overline\beta/\lambda$&7.54&8.02\\ \hline\hline
$h$&.738&.803\\ \hline
$k$&.337&.315\\ \hline
$l$&.580&.542\\ \hline\hline
$\alpha$&6.54&7.02\\ \hline
$\nu_0$&$-11.0$&$-9.56$\\ \hline
$\nu_1$&.0174&.0201\\ \hline
$\nu_2$&$2.4\times10^{-7}$&$2.3\times10^{-7}$\\ \hline
\end{tabular}
}
\caption{\label{tab1} Parameter estimates. Values (in adequate powers of weeks) are based on the optimal choice of $q=p_2\eta$ (cf.~Fig.~\ref{fig6}AB) and the most likely duration of cross-protection (Fig.~\ref{fig7}). The computation uses steady state solutions of \eqref{for032} and \eqref{for032EkI}, where $\beta$ is approximated by the mean value.
}
\end{table}

The other parameters ($h,k,l$; cf.~Table~\ref{tab1})
are calculated with the help of steady state solutions ($S'=E'=I'=R'=0$) and mass-action \eqref{for004}: from \eqref{for032EkI} we extract expressions for $E$ and $I$, plug them into the steady state versions of \eqref{for032}, and solve for $N'$.
The result maps the mean duration of cross-protection $1/\rho$ to the time-derivative of the population size $N'$; first order expansion offers a good approximation, 
\begin{align}\label{for034}
 N'(1/\rho)=\nu_0+\nu_1\left(1+\alpha/\kappa\right)/\rho+\underbrace{\nu_2\left(1+\alpha/\kappa\right)^2/\rho^2+\dots}_{\approx0\,,~\text{cf. Table \ref{tab1}}}\,.
\end{align}
The formula contains the transmission rate ratio $\kappa=k/l$.
For our model data, where $N'=0$, this ratio can be calculated: $\kappa_0=.581$.
The situation is different for real data from Thailand, as these data involve demographic changes \cite{CIL08} (i.e., $N'\neq0$). Nevertheless, we use $\kappa=\kappa_0$ as an estimate; this value is close to the one ($\kappa=1/1.85$) mentioned in  \cite{CIL08} (legend to Fig.~6A).

\begin{figure}
\center\includegraphics[height=49ex]{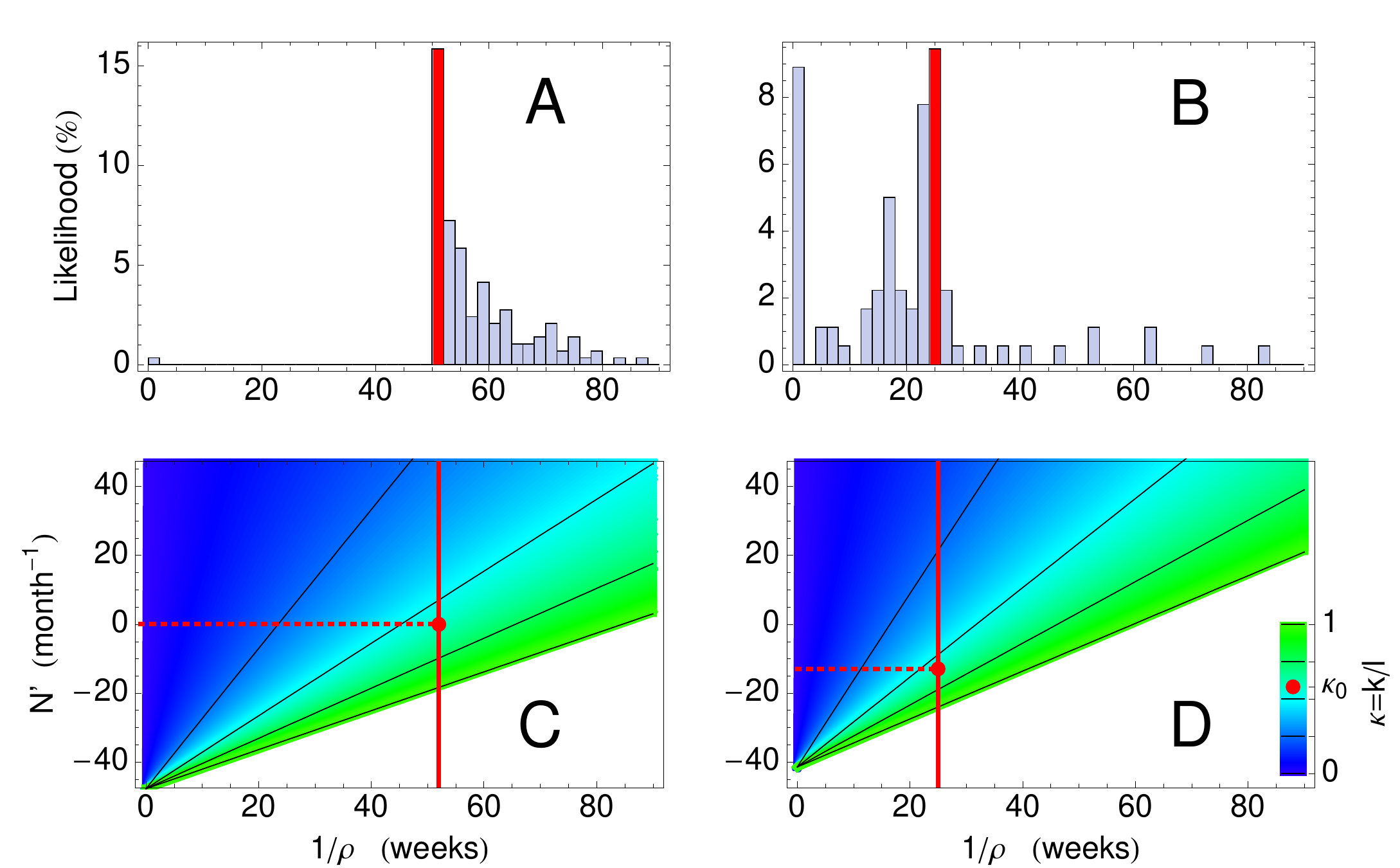}
\caption{\label{fig7}Clinical cross-protection versus demographic transitions. Similar to Figure \ref{fig6}, the diagrams on the right hand side represent dengue data from Thailand, the ones on the left hand side the model scenario in \cite{NK08}.
Panels A and B show the likelihoods for the various mean durations of cross-protection --- obtained by shifting the time line underlying the reconstruction and minimizing the squared distance between the time-shifted reconstruction and the original (blue) beta curve (in Fig.~\ref{fig6}).
The size of the time-shift represents the duration of cross-protection, the frequency of the distance minimums measures the corresponding likelihood.
Panels C and D illustrate the (nearly linear) relationship between the mean duration of cross-protection $1/\rho$ and the change of the effective population $N'$
with respect to the transmission rate ratio $\kappa=k/l$ (blue-greenish color).
The model data incorporate cross-protection for a period of one year (cf.~legend to Fig.~4 in \cite{NK08}), corresponding to the likelihood maximum (red) in Panel A. For real data (Panel B), the most likely period of cross-protection is slightly below half a year.
When utilizing the transmission rate ratio from the model data (red dot), the Thailand data show a declining effective population ($N'=-12.9/\text{month}$) as proposed in \cite{CIL08}.}
\end{figure}

As the parameter $\rho$ does not enter the truncated dynamics \eqref{for032EkI} explicitly, we model clinical cross-protection
by shifting the time line and estimate its duration by multiple reconstructions.
We calculate likelihoods for all possible durations
(Fig.~\ref{fig7}AB).
Our model data show a peak at one year (Fig.~\ref{fig7}A), which is in perfect agreement with the simulation setting in \cite{NK08}.
Real data from Thailand show several peaks --- the latest and highest one at half a year (Fig.~\ref{fig7}B). A half-year period corresponds to a decreasing effective population (cf.~the dashed line in Fig.~\ref{fig7}D) and thus confirms the conclusions in \cite{CIL08}.

Therefore, our results confirm that clinical cross-protection and demographic transitions --- suggested to explain the discrepancy between DHF counts and decreasing transmission rates --- represent possible and non-exclusive mechanisms.
Without many changes we were able to extend our method to an infection that involves several stages. We learned that the method is relying on an SIR-structure similar to the one of the measles example but with exposed individuals taking over the role of susceptibles.
Multiple application of the reconstruction procedure, as required in our third example, will represent another extension.

\subsection{Influenza}

\begin{figure}
\center\includegraphics[height=150mm]{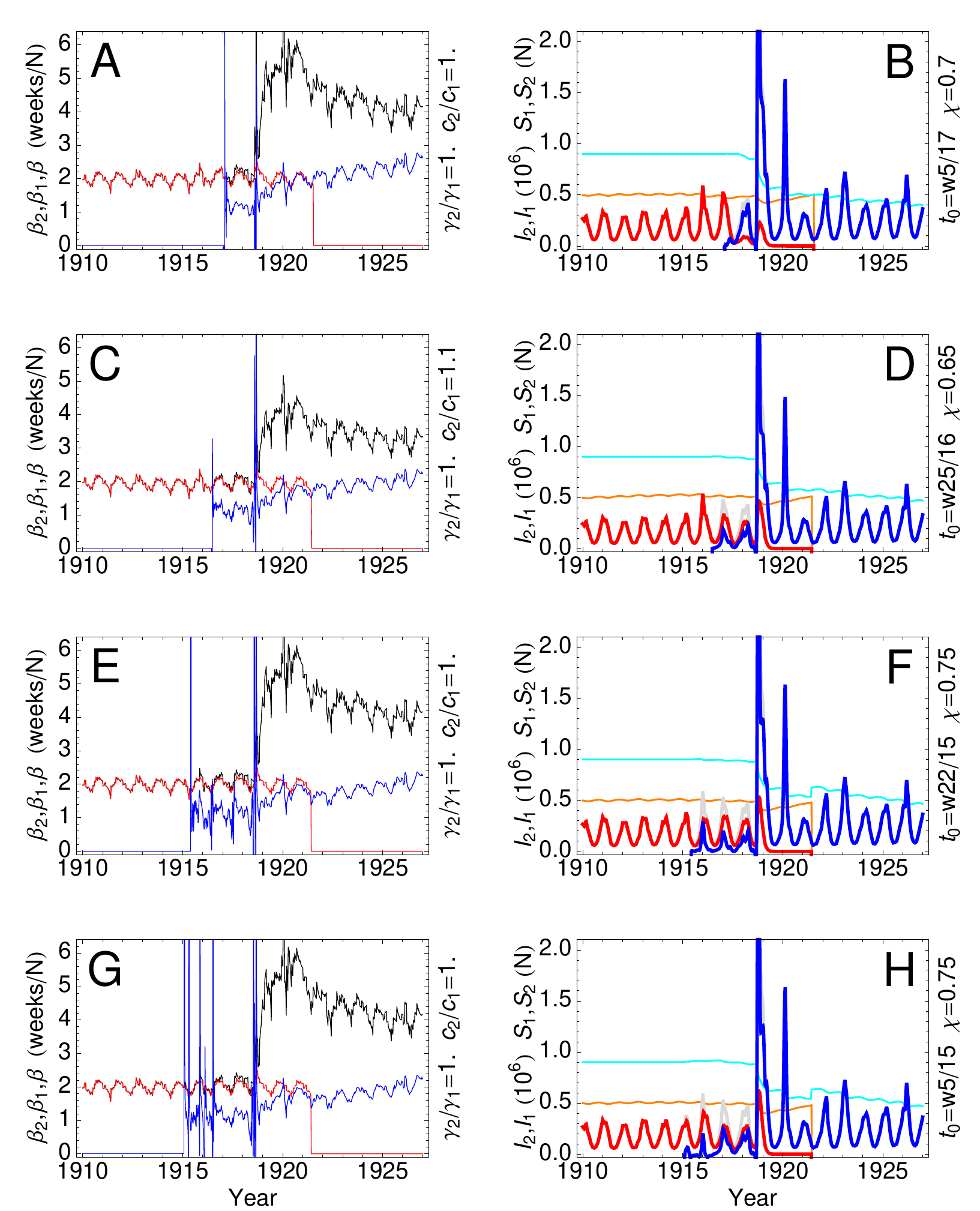}
\caption{\label{fig8} Exchange of strains.
The four sets of diagrams (Panels AB--GH) illustrate invasion scenarios, which  differ by the time $t_0$ the pandemic strain is introduced into the human host population (cf.~figure labels).
The diagrams on the right hand side show the time evolution of infectives and susceptibles, $I_1$ (red), $I_2$ (blue), $I_1+I_2$ (light gray), $S_1$ (orange), $S_2$ (cyan) --- the diagrams on the left hand side the corresponding transmission rates, $\beta_1$ (red), $\beta_2$ (blue), $\beta$ (gray).
Infectious periods of $1\,\text{week}=1/\gamma_1=1/\gamma_2$ ensure that the scale ``weeks/N'' corresponds to units of $\R_0$. The other parameter values ($N=1.5\times10^8$, $c_1=1\%$, $\R_0=2$) are chosen so that the moving average of $\beta_1(t)$ is constant before the pandemic peak $t_p$. These choices define a realistic model scenario \cite{MRL04}.
The transmission rate of the resident strain $\beta_1(t)$ ($=\beta(t)$ for $t<t_0$) is continued beyond $t_0$ by a combination of the annual shape $\overline {\beta}(t)$ prior to the peak and a copy of $\beta(t)$ that is stopped right after the peak, $\beta_1=(1-r)\overline\beta+r\beta$.
Ratios $r>0$ support the assumption that $I_1$ contributes to the peak; here we chose $r=1/4$.
Independent of this choice, $I_1(t)$ quickly declines after the peak. At values below $1$ (Panels B,D) and below $10^{-2}$ (Panels F,H) the resident strain is removed from the system.}
\end{figure}

We apply the method to multiple pathogenic strains, examining mortality data \cite{IIDDA} related to the 1918/19 influenza pandemic in the U.S.
Here we try to understand how the pandemic strain has replaced the residing strain after entering the host population.
To this end, we determine the strain-specific infections for periods covering the pandemic peak and reconstruct the corresponding transmission rates.
Case-fatality ratios enable us to express the death rate as a linear combination over the strains,
\begin{align}\label{for028}
M=\lambda_1I_1/c_1+\lambda_2I_2/c_2\,.
\end{align}
Transmission rates do not superpose in a linear fashion,
\begin{align}\label{for031}
\beta\neq\beta_1+\beta_2\,,
\end{align}
which makes the reconstruction problem a non-trivial task.

Before we start, we like to emphasize that our modeling approach is again minimalistic.
Unlike \cite{HDDME13}, for example, we do not incorporate extra information (about school terms, temperature changes, or changes in human behavior) in addition to the mortality data around 1918/19.
We only propose two standard SIR models --- one for the residing strain (1) and one for the invading (supposedly pandemic) strain (2).

The time evolution of the SI-compartments of the residing strain is given by
\begin{subequations}\label{for025}
\begin{align}
S_1'&=\rho_1(N-S_1-I_1)-\Delta_2 S_1-\beta_1S_1I_1/N\\
I_1'&=\beta_1S_1I_1/N-\lambda_1I_1\,.
\end{align}
Replenishment of susceptibles is dominated by the decay of immunity, $\rho_1\approx\delta$ (with $1/\delta=5$--$10\,\text{years}$, as common for flu).
The equations include two further assumptions.
\emph{First.} 
Previously infected individuals are partly cross-protected against the other strain. That is, the invading strain, which infects $S_2(t_0)-S_2(t)=-\int_{t_0}^tS'_2(\tau)\,d\tau$ individuals up to time $t>t_0$, reduces the number of individuals susceptible to the residing strain approximately at the rate
\begin{align}\label{for026}
\Delta_2 S_1=\chi\frac{S_1}{N}S_2'\,.
\end{align}
\end{subequations}
The parameter $\chi\le1$ defines the amount of cross-protection.
\emph{Second.} We reconstruct the transmission rate $\beta_1(t)$ from data prior to the invasion, when only strain 1 is present, and --- assuming that strain 1 does not change phenotypically --- extrapolate its averaged annual shape (red curves on the left-hand side panels in Figure \ref{fig8}) beyond the pandemic peak.

Motivated by the neutral network mechanism \cite{KCG06}, the invading strain is modeled by an almost identical copy of the residing strain \eqref{for025}, 
\begin{subequations}
\begin{align}\label{for027a}
S_2'&=\rho_2 (N-S_2-I_2)-\Delta_1S_2-\beta_2S_2I_2/N\\\label{for027b}
I_2'&=\beta_2S_2I_2/N-\lambda_2I_2\,.
\end{align}
\end{subequations}
Initially, susceptibles to the invading strain $S_2$ are only replenished by birth, $\rho_2\approx\mu\ll\delta$.
That is, the replacement of $S_2$ is small.
Therefore, \eqref{for027a} is approximated by $S_2'=-\Delta_1 S_2-\lambda_2I_2-I_2'$, where due to \eqref{for027b} the contact term $\beta_2S_2I_2/N$ is expressed by $\lambda I_2+I_2'$.
According to \eqref{for028}, we further replace $I_2$ and $I_2'$ by $M$ and $I_1$.

The number of susceptibles to the invading strain $S_2$ changes around the pandemic peak $t_p$. Before and after the peak, $S_2(t)$ can be approximated by a constant. That is, the cross-protection term reads 
\begin{align}\label{for048}
\Delta_1 S_2(t)=\chi\left(\frac{\theta(t-t_p)}{\R_0^1}+ \frac{\theta(t_p-t)}{\R_0^2}\right)S_1'(t)\,,
\end{align}
where $\theta(t)$ denotes the unit step function and $\R_0^k$ an estimate for the basic reproduction of strain $k$.

The number of susceptibles to the residing strain $S_1(t)$ is nearly constant, even in the vicinity of the pandemic peak. That is, $S_1'(t)\approx0$. One may therefore neglect cross-protection for the invading strain \eqref{for048} and only consider contacts in \eqref{for027a}. The outcome barely changes.

Either way, $S_2'$ is defined in terms of $M$, $M'$, $I_1$, $I_1'$, and $S_1'$. As cross-protection $\Delta_2S_1[S_1,S_2']$ is the only remaining term containing the index 2, the system of strain 1 \eqref{for025} can now be solved.
Finally, the transmission rate of the invading strain is calculated by
\begin{align}\label{for030}
\beta_2&=\frac{I_2'+\lambda_2I_2}{S_2I_2}\,N\,,
\end{align}
where again $S_2$, $I_2$ are expressed by $M$, $I_1$ and their derivatives.

Following the instructions outlined above we perform numerical simulations where we vary the time $t_0<1918$ at which the invading strain is introduced into the population.
The blue curves in the left-hand side panels of Figure \ref{fig8} show the evolving transmission rates $\beta_2(t)$ in four example scenarios.
Although these scenarios vary largely with respect to the choice of $t_0$, the beta reconstruction produces very similar results.
For up to three years after introduction, the transmission rate of the invading strain ($I_2\gg0$) fluctuates around values of 1 (in units of $\R_0$) until suddenly it becomes singular, resembling a Dirac delta at the pandemic peak $t_p$. A closer look reveals that $\beta(t_p)$ never exceeds values 5--7 (in units of $\R_0$). Then $\beta_2(t)$ increases and approaches the values of the residing strain $\beta_1(t)$ (red) as $t\to\infty$.

The gray curves in the left-hand side diagrams of Figure \ref{fig8}
illustrate the reconstruction performed without differentiation between the two strains. One observes a huge increase of $\beta(t)$ at the pandemic peak, which only stops after a couple of years.
We confirm that $\beta$ is not the result of a linear superposition of $\beta_1$ and $\beta_2$.
Except for the singularity at the pandemic peak, the transmission rates of the invading strain $\beta_2(t)$ behave very regularly and do not seem to resemble a pandemic at all. 
There is no need to model a higher-than-normal case-fatality ratio \cite{TM06} and/or a shorter infectious period for strain 2. In reality, parameters might have changed back to normal values (characteristic for annual flu) very quickly, explaining why small (or no) changes of these parameters accomplish a valid effective theory (cf.~Fig.~\ref{fig8}CD).

The scenario in Figure \ref{fig8}G, where the pandemic strain is introduced at the earliest time $t_0$ ($\text{week}~5/1915$), shows strong fluctuations with singularities.
Even more irregular behavior of $\beta_2(t)$ occurs in scenarios of yet earlier $t_0$, although certain parameter combinations ($c_2$, $\chi$) produce singularities also if strain 2 is introduced later.
Scenarios with singularities of $\beta_2(t)$ are unrealistic.
The singularity at $t_p$ is an exception, if interpreted as the initial cause of the pandemic. We cannot decide if changes of the virus (mutation or reassortment) are responsible for the singularity of $\beta_2(t)$ in 1918.
Our computations only suggest that the pandemic could have been initiated by
any kind of singularity in the transmission rates.
Sudden changes in the host-contact behavior, for example, induced by extended world-wide migration at the end of the war, represent another plausible cause.

Herald waves could have started as early as 1915/16. The reconstruction scenarios in Figure~\ref{fig8} indicate the possibility of three epidemic waves prior to the pandemic peak --- periods during which both strains could have been present. The proportion of the two strains depends on our model parameters (e.g., $\chi$) but in no obvious way. Even if the prevalence of the invading strain is far above zero, the corresponding transmission rate prior to the pandemic peak never exceeds 1 much. However, the beta reconstruction always predicts an exchange of strains --- at or immediately after the pandemic peak, where the prevalence of the residing strain reduces to nearly vanishing amounts.

The precise dynamics underlying the exchange of strains around 1918/19 was certainly more complicated.
Nevertheless, the similarity of our results (Fig.~\ref{fig8}, right) to those in \cite{KCG06} (Fig.~3A) demonstrates that the modeling approach via the reconstruction of transmission rates can unveil features inherent to the antigenic shift of influenza.

\section{Discussion}

We have developed a methodology that enables us to formulate models of disease transmission by extracting information from time series of incidence or mortality.
Based on the assumption that the epidemic is governed by SIR dynamics
(i.e., infection related compartments and some version of mass action) we 
reconstruct transmission rates as a function of time.

The reconstruction is given by explicit formulas
(\ref{for00252a},\,\ref{for00252b}),
with data represented by functions defined on a continuous time scale.
Mathematically these formulas are easy to handle and straightforwardly provide us with error estimates \eqref{for0100}.
In numerical computations we replace the continuous-time expressions by carefully chosen discrete versions --- critically important for the integrals involved; see \eqref{for049}.
Doing so, we are able to reconstruct transmission rates for data of measles, dengue, and influenza.
These examples, which cover very different types of infections \cite{LF09}, illustrate the method's wide range of application.

Besides mathematical and conceptual flexibility, the involved computations are fast. The core expression is a convolution integral, which determines the time complexity of the reconstruction to be quadratic.
It takes less than two minutes to compute Figure \ref{fig2}B on a usual laptop. Using a probabilistic sampling method (e.g., \cite{MF05,CF08}), it may take a day to accomplish the reconstruction for a data set of similar size; 20 hrs is the reference given in \cite{CF08}.

The large difference in computation time is easy to understand.
Sampling methods offer probabilistic analysis with confidence intervals whereas our method only determines the error resulting from error estimates of the input variables.
Moreover, sampling provides intrinsic parameter estimates, which a priori our approach is lacking.
Therefore, the comparison is not entirely fair.
The more recent methods \cite{HERE11,WCBIL12,WYCIL13}, which utilize nonlinear optimization techniques, are more similar to our compartmental approach. 
Yet, determining beta values and parameters at once, these methods need much more computing power and, concluding from the parameter behavior, might not gain so much.

Namely, when performing the reconstruction we learned that most of the parameters --- removal rates and initial values --- are not required to be precise. Different infectious periods, for example, produce similar outcomes, at least on a large time scale (cf.~Figs.~\ref{fig2}EF, \ref{fig4}E-H). Mortality is sufficiently defined by mean values (Fig.~\ref{fig1}B).
Initial values can be estimated using steady state solutions and external data. Their influence decreases exponentially --- with decay of immunity or inverse lifetime --- so that that asymptotically, for large time, the reconstruction returns the correct transmission rates.

Only one parameter must be taken from a small range of values as otherwise the reconstruction yields unrealistic results or even collapses (Fig.~\ref{fig2}C). For measles and flu, this parameter is the effective population size (if assumed to be of constant size), for dengue, it is the DHF-likelihood of secondary infections.
Suitable parameter values are determined by the long-term behavior of beta.
That is, a posteriori our method provides necessary parameter estimates.

Previous work on pre-vaccination data of measles infections in the U.K. \cite{FC82,FG00,BFG02,MF05,CF08} offered a successful test of the method. We reconstructed the expected annual patterns of high and low transmission.
More importantly, the reconstructed betas reproduced the original data. 
For dengue, we immediately re-obtained the data (Fig.~\ref{fig6}D) from the reconstructed betas (Fig.~\ref{fig6}F). For measles, reproduction did only work for a few years (Fig.~\ref{fig4}E). Here we had to increase the time-resolution, which turned out to have a huge impact on the error of the computations (cf.~Fig.~\ref{fig4}AC).
After interpolation, the error of the reconstructed betas disappeared (Figs.~\ref{fig4}BD). With those betas we were then able to correctly reproduce the original data (Fig.~\ref{fig4}F-H).

We also tested a simplified version of the method,
which proposes an effective population of constant size ---
useful if demographic data are not available.
The reconstruction results turned out to be quite similar to those based on a variable population size (Fig.~\ref{fig3}).

Elaborate probabilistic simulations are often difficult to verify.
Here our reconstruction can serve as a tool, allowing to cross-check assumptions in a simple SIR setting.
The dengue application is an example where the method was used to re-evaluate two seemingly contradictory hypotheses, which claim to explain controversial observations related to the re-emerging dengue epidemic in Thailand \cite{TNS08}.
Our reconstruction has shown that the two hypotheses, the one based on a medical \cite{NK08} and the other on a demographic mechanism \cite{CIL08},
are both plausible and not contradicting.

The beta reconstruction can be combined with other methods and approaches, leading to novel hypotheses relying on the time evolution of transmission rates. This has been demonstrated here with the influenza example.
Having to differentiate between multiple strains \cite{SLJ04,KCG06}, flu was the most advanced application out of the three.
We performed the reconstruction based on mortality data and simultaneously modeled the evolution of an invading strain. The resulting transmission rates suggest that the invasion could have happened many month ahead of the pandemic (Fig.~\ref{fig8}), which is telling us about the origin of herald waves.

\section*{Acknowledgements}

This work was supported in part by a grant to David Earn, Troy Day, Jonathan Dushoff, and Junling Ma from the Canadian Institutes of Health Research (CIHR). 
The author thanks these researchers for valuable discussions during his stay at McMaster University.

\end{document}